\documentclass{article}
\usepackage[margin=1in]{geometry}
\usepackage[utf8]{inputenc}
\usepackage{fancyhdr}
\usepackage[pdftex,bookmarks=true]{hyperref}
\usepackage{natbib}
\usepackage{aas_macros}
\usepackage{wrapfig}
\usepackage{graphicx}
\usepackage{xcolor}
\usepackage{amssymb}  
\usepackage{amsmath}  
\usepackage{multicol}
\usepackage{placeins}

\begin{document}
\pagestyle{fancy}
\pagenumbering{arabic}

\lhead{Short-Period Planets and Planet Formation/Evolution}
\chead{}
\rhead{\thepage}	
\lfoot{}
\cfoot{}
\rfoot{}
\setcounter{page}{0}

\newcommand{\tess}{\emph{TESS}}
\newcommand{\kepler}{\emph{Kepler}}
\newcommand{\corot}{\emph{CoRoT}}
\newcommand{\ktwo}{\emph{K2}}

\thispagestyle{empty}
\begingroup 
 \centering
 \large Investigating Planet Formation and Evolutionary Processes with Short-Period Exoplanets\\
\endgroup

\vspace{12 pt}

\noindent {\bf Primary Author:} Brian Jackson -- \href{mailto:bjackson@boisestate.edu}{bjackson@boisestate.edu}\\
\indent Boise State University, 1910 University Drive, Boise, Idaho 83725-1570\\
\indent (208) 426-3723\\

\noindent {\bf Co-Authors:} Elisabeth Adams (\href{mailto:adams@psi.edu}{adams@psi.edu} - Planetary Science Institute), Ren\'e Heller (\href{mailto:heller@mps.mpg.de}{heller@mps.mpg.de} - Max Planck Institute for Solar System Research, Germany), \& Michael Endl (\href{mailto:mike@astro.as.utexas.edu}{mike@astro.as.utexas.edu} - UT Austin/McDonald Observatory)


\paragraph{Abstract.} From wispy gas giants on the verge of disruption to tiny rocky bodies already falling apart, short-period exoplanets pose a severe puzzle to theories of planet formation and orbital evolution. By far most of the planets known beyond the solar system orbit their stars in much tighter orbits than the most close-in planet in the solar system, Mercury. Short-period planets experienced dynamical and evolutions histories distinct from their farther-out cousins, and so it's not clear they are representative of all planets. These exoplanets typically have radii between about 1 and 4 Earth radii, whereas the solar system does not contain any planet in this radius range. And while the most massive planets in the solar system occupy the icy regions beyond about 5\,AU from the sun, about 1\,\% of sun-like stars have a Jupiter-mass planet near 0.05\,AU, with just a few days of an orbital period. How did these short-period planets get there? Did they form in-situ, or did they migrate towards their contemporary orbits? If they migrated, what prevented them from falling into their stars? Vice versa, could some of the remaining 99\,\% of stars without such a hot Jupiter show evidence of their past consumption of a close-in, massive planet? The proximity between short-period planets and their host stars naturally facilitates observational studies, and so short-period planets dominate our observational constraints on planetary composition, internal structure, meteorology, and more. This white paper discusses the unique advantages of short-period planets for the theoretical and observational investigations of exoplanets in general and of their host stars.



\newpage
\begin{figure}
\includegraphics[width=\textwidth]{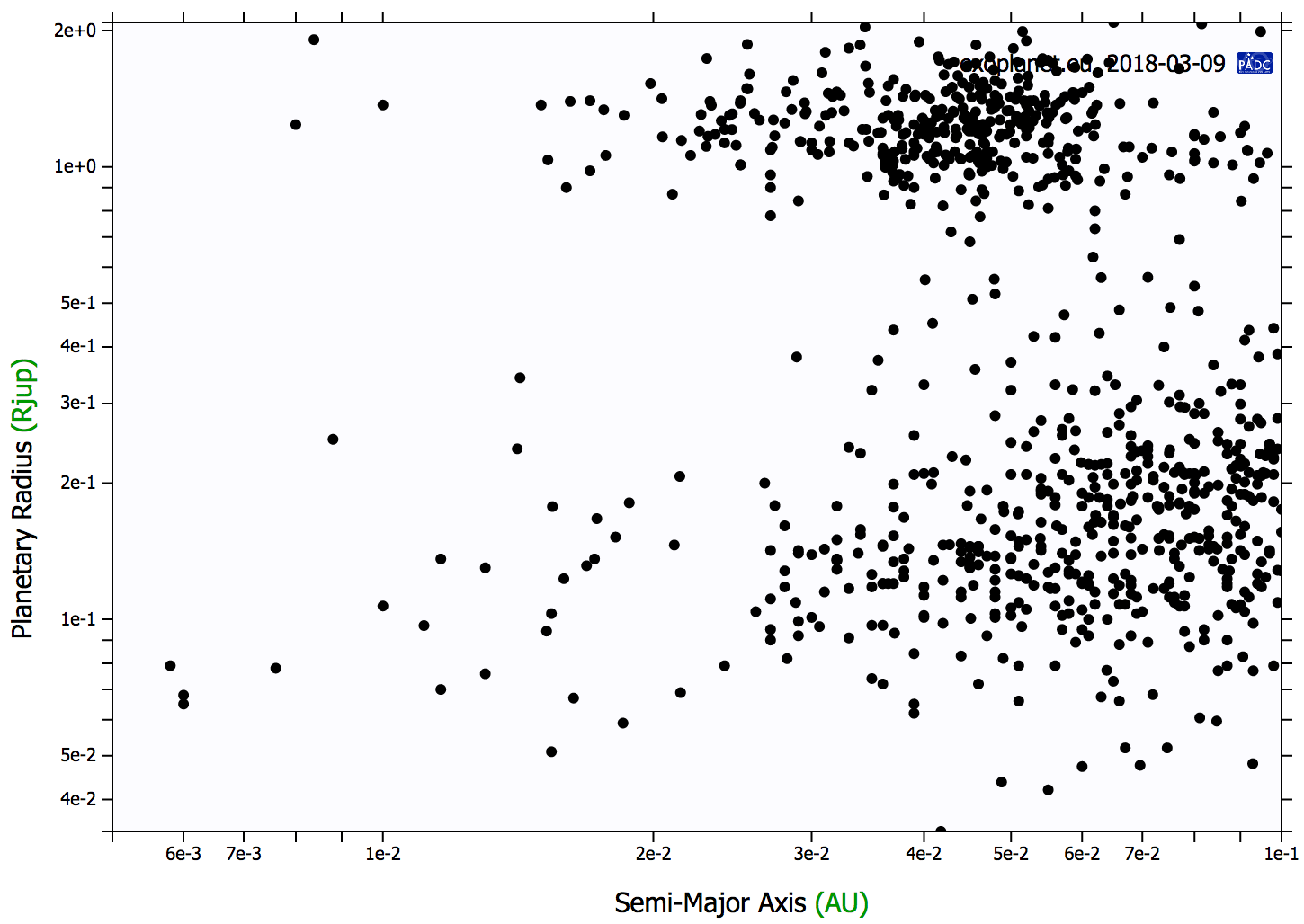}
\caption{Planetary radii in Jupiter radii $R_{\rm Jup}$ vs. semi-major axes in AU for confirmed and candidate short-period planets, which include ultra-short-period planets (USPs) with periods less than 1 day. \label{fig:USPs_Rp-vs-a}}
\end{figure}

The discoveries of almost two hundred bodies with orbital periods $\lesssim~1$ day (ultra-short-period planets, USPs -- \citealp{2013ApJ...774...54S, 2013ApJ...779..165J, 2014ApJ...787...47S, 2016AJ....152...47A, 2017AJ....153...82A}) and of hundreds of Jupiter-sized planets with orbital semi-major axes $\lesssim$ 0.1\,AU continues the surprising trend that planetary systems throughout the galaxy may diverge dramatically from those in the solar system. According to canonical planet formation theory, many of these short-period planets, particularly the USPs, cannot have formed where we find them today. Thus, short-period planets and USPs provide a critical test for theories of planet formation and evolution.

Although they challenge our understanding, the very short periods substantially facilitate detailed follow-up and characterization, and planned and upcoming missions and surveys promise to reveal additional clues to these planets' origins and natures. The ineluctable bias in transit surveys toward short periods means many of the smallest planets uncovered by the TESS Mission will be short-period. Large-scale surveys such as TESS, LSST, and PLATO may provide hints about the fates of hot Jupiters. Given their observational prominence, it is important to understand whether short-period planets are representative of all exoplanets, particularly the smallest and least massive planets for which follow-up is especially difficult, or whether they constitute a distinct species. In this white paper, we discuss aspects of our current understanding of short-period planets and explore possible avenues for future research. 

\newpage
\section{Theoretical Considerations}
Current thinking includes three origin scenarios for short-period exoplanets: (1) they formed in-situ \citep{2000Icar..143....2B,2016ApJ...829..114B}; or they formed farther out and (2) were gravitationally excited into highly eccentric (and potentially inclined) orbits, and tidal interactions with their host stars circularized and shrank the orbits \citep[e.g.][]{2008ApJ...686..580C}, or (3) they were brought in by gas disk migration on (possibly) low-inclination orbits \citep[e.g.][]{1996Natur.380..606L}. Determining which of these scenarios applies to short-period planetary systems requires a number of detailed, astrophysical studies, as we describe below.

\subsection{Gas Disk Migration Stopping Mechanisms}
\begin{wrapfigure}{l}{0.425\textwidth}
\includegraphics[width=0.425\textwidth]{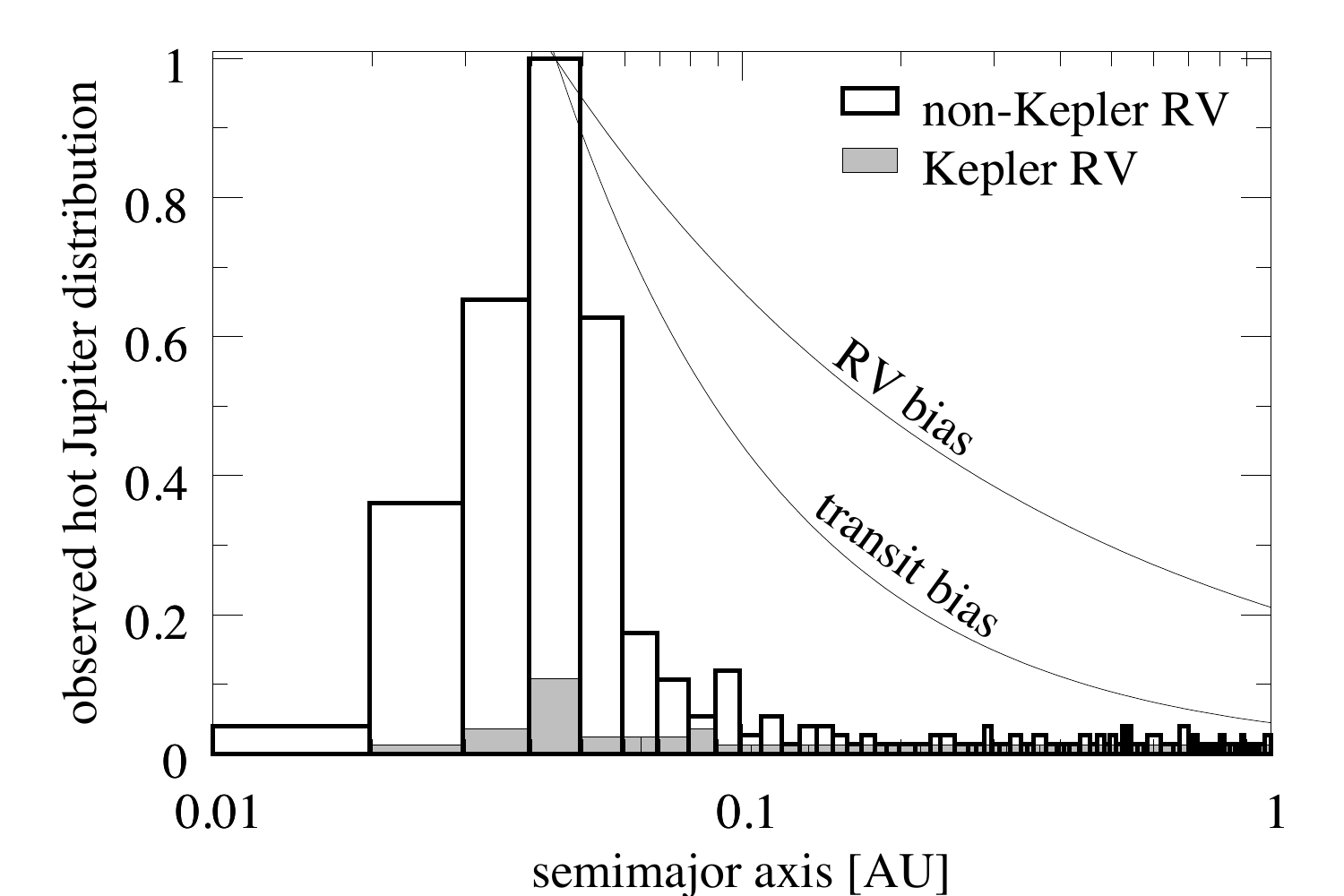}
\caption{Distributions of semi-major axes $a$ for planets with masses $>0.1\,M_{\rm Jup}$ around Sun-like stars, found either by the Kepler mission or another survey. The solid lines illustrate the radial velocity amplitude and the geometric transit probability, respectively. Figure from \citet{Heller2018} with data from Exoplanet Archive (10 Jan 2018).} \label{fig:exoplanets}
\end{wrapfigure}
For hot Jupiters, none of the above scenarios satisfies all constraints \citet{2018arXiv180106117D}. Instead, there are probably contributions from both (2) and (3) (with [1] possibly less important). For (2), a cogent picture is emerging that can explain the observed wide distribution of stellar obliquities \citep{2012ApJ...757...18A} but has trouble accounting for Jupiter-like bodies more distant from their host stars, warm Jupiters. For (3), recent work on disk misalignment mechanisms can account for the stellar obliquity distribution \citep{2015ApJ...811...82S}, but what stops the inward migration is unclear. 

Competing theories to halt disk migration include clearing of the close-in protoplanetary disk via magnetorotational instabilities \citep{2002ApJ...574L..87K}, planet-disk magnetic interactions \citep{2003MNRAS.341.1157T}, and tidal halting \citep{1998ApJ...500..428T}. Although the latter mechanism exhibits good agreement with observations \citep{2013ApJ...769...86P}, there remains no consensus of how to explain (1) the sharp pile-up of hot Jupiters at 0.05\,AU around Sun-like stars (Fig.~\ref{fig:exoplanets}); (2) the absence of a hot Jupiter pile-up in sample from Kepler  \citep{2012ApJS..201...15H}; (3) the long-term stability of hot Jupiters under tidal dissipation in the star; and (4) a pile-up of planets in the subsample of Kepler planets with RV measurements (Fig.~\ref{fig:exoplanets}). 

\citet{Heller2018} argues that the pile-up is a result of competing torques from the stellar tides, which act to repel a close-in planet beyond the co-rotation period of $\sim~1$\,d, and the disk torque, which drives planet migration towards the star. The torques of highly dissipative, fast-spinning stars with protoplanetary disks of low gas surface densities could produce a zero torque at the observed pile-up period (or semi-major axis). In this scenario, hot Jupiters trace the early spin history of their host stars and even the properties of the long-gone proto-planetary disks in which they formed.

\subsection{Disruption of Hot Jupiters}
Under whatever scenario hot Jupiters achieved orbits that passed near to their host stars, tidal interactions subsequently shaped their orbits. Under scenario (2) above, tidal dissipation heated the planets' interiors, circularizing and shrinking the orbits \citep{2007ApJ...669.1298F, 2008ApJ...678.1396J, 2015ApJ...798...66D}. Once the orbits were circular and the planets' rotations had become tidally locked, the planets could still raise large enough tides on the host stars that tidal interactions could modify the stellar obliquities \citep{2010ApJ...718L.145W} and drive orbital decay, possibly resulting in disruption via Roche-lobe overflow and/or planetary accretion \citep{2015ApJ...813..101V, 2017ApJ...835..145J}. 

The speed and frequency with which hot Jupiters may undergo disruption and accretion is unclear and likely depends on the efficiency with which tidal dissipation operates within stars (i.e., the stellar tidal $Q$-value), what happens to the gas once it escapes the planet \citep{2017MNRAS.465..149J}, and the mass-radius relationship \citep{2017ApJ...835..145J}. Thus, detailed modeling efforts of star-planet tidal interaction, possibly coupled with stellar evolution models \citep{2016CeMDA.126..275B,Heller2018}, are required to determine the fates of hot Jupiters. 

\subsection{USPs As Possible Signposts for Other Planets}
How origin scenarios (1), (2), and (3) above apply to USPs is unclear. In-situ formation seems less likely, given that protoplanetary disks may not extend inward to periods of hours \citep{2007prpl.conf..479B}, and in such close orbits, even the most refractory solids could not condense \citep{2011ApJ...740..118L, 2003ApJ...591.1220L}. Regarding (2), \citet{2006ApJ...638L..45F} showed the original pericenters for the highly eccentric orbits were about half the semi-major axis for the final, circularized orbit, so the latter distance must be at least twice a planet's Roche limit. However, many USPs orbit within twice their Roche limit \citep{2017ApJ...835..145J}. Scenario (3) coupled to subsequent orbital decay may explain the origins of USPs massive enough to raise substantial tides on their host stars (several Earth masses) \citep{2017ApJ...842...40L}, but less massive USPs may require resonant/secular interactions with other bodies, likely other planets, coupled to tidal dissipation within the USP, to drive orbital decay, as in the 55 Cnc system \citep{2015MNRAS.450.4505H}.

This latter hypothesis makes the prediction that the presence of a USP requires additional, nearby planets. As a consequence, USPs could be signposts of yet undiscovered planets in a system. Indeed, considering the population of USPs known from the Kepler mission, \citet{2014ApJ...787...47S} found that the frequency with which additional transiting candidates were observed in USP systems was consistent with their all being multi-planet systems. To the viability of this hypothesis, dynamical analyses and $N$-body orbital simulations of known multi-planet USP systems are required.

\section{Observational Prospects}
\subsection{Mass Measurements}
\begin{figure}
\includegraphics[width=\textwidth]{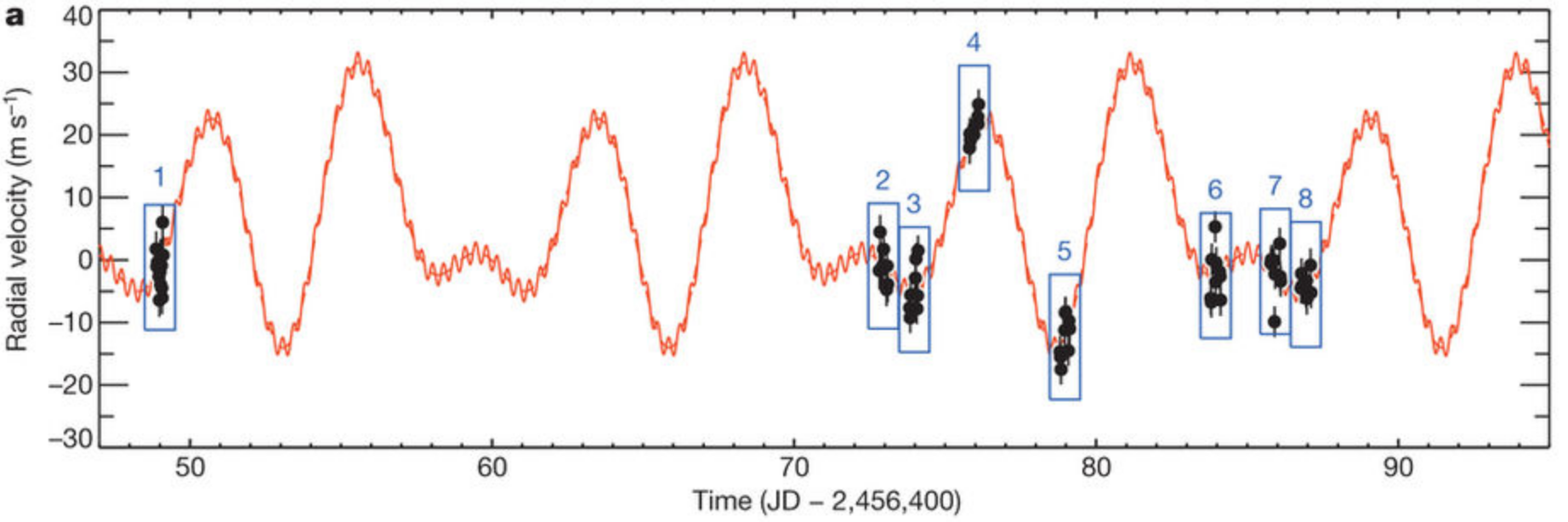}
\caption{RV measurements for Kepler-78 from \citet{2013Natur.503..381H} (black dots). The modeled RV signal (orange curve) includes a large-amplitude component of stellar jitter on a time scale of days and the planet's 8-hour RV signal with a much smaller amplitude. \label{fig:Howard2013_Fig1a}}
\end{figure}

Now that radial velocity (RV) surveys are regularly breaching the $1\,{\rm m\,s}^{-1}$ barrier of systematic noise \citep{2011A&A...525A.140D}, stellar jitter has become the main limitation to our ability to detect Earth-mass planets. With orbital periods as short as several hours, small USPs induce larger RV signals than longer-period planets of similar masses. Equally important, planet-induced orbital RV variations on time scales of hours are often different from the time scales of stellar activity. And finally, fewer observations are required to recover a USP's RV signal since multiple orbits can be observed in very little time. For example, \citet{2013Natur.503..381H} successfully recovered Kepler-78 b's RV signal over the course of just a few nights, even though the star exhibited RV variability many times larger than the planet's signal (see Figure \ref{fig:Howard2013_Fig1a}).

\subsection{Signals of Planetary Disruption and Accretion}
Estimates of the frequency and violence of hot Jupiter disruption and accretion could inform observational searches for these processes as the consumption of a planet by its host star likely produces distinctive signals in time-series data. \citet{2012MNRAS.425.2778M} estimates that, in the weeks prior to a short-period planet's final plunge into its star, the system's peak luminosity in UV and X-rays may exceed $10^{36}$ erg s$^{-1}$, after which the stellar surface is enshrouded by an outflow driven by the merger, followed by an optical transient at a luminosity of $10^{37}$-$10^{38}$ erg s$^{-1}$ on a time-scale of days. Upcoming and previous long-baseline optical surveys such as Kepler/K2, TESS, LSST, and PLATO may provide datasets ideally suited for seeking these merger signals.

\subsection{TESS Yield of USPs}
\citet{2014ApJ...787...47S} estimated that something like 1 in 200 Sun-like stars host USPs, making them relatively rare among the species of exoplanet. The Kepler mission found hundreds of USPs, and given TESS's larger number of observed stars, many of which will be observed with high time-resolution (2-min), this mission is likely to find many more USPs. To estimate the number of USPs that TESS will find, \citet{2015ApJ...809...77S} used the Kepler occurrence rates and applied it to an approximated sample of stars that would make up the TESS catalog, then drew a sample of objects. Using their Table 6, we find that TESS is estimated to discover 104 USPs with periods between 0.5 days $\le P \le$ 1 day. However, because the Kepler data upon which they based these estimates did not include even shorter period planets with $P < 0.5$ d, these planets are not included in the catalog. Among all known USPs as of March 2018, about 10\% have $P < 0.5$ days.

Of the Sullivan sample of 104 USPs that TESS should find: (1) many are small -- just 2 are hot Jupiters and the rest have $< 3\ R_{\rm Earth}$. (2) many orbit faint stars, but 10 are $V < 10$ or brighter and another 11 are $10 < V < 13$; (3) many have friends -- only 8 have no companion planets (9 have 1, 21 have 2, 50 have 3, and 16 have 4 other planets); and (4) their stars tend to be small -- there are 88 M dwarfs, 7 K, 6 G, and just 3 around stars larger than 1.2 $R_{\rm Sun}$. As such, many of their masses could be measurable with the current RV instruments: the median RV signal is about 5 m/s.

\subsection{Obliquities of Short-Period Planets}
\begin{wrapfigure}{l}{0.5\textwidth}
\includegraphics[width=0.5\textwidth]{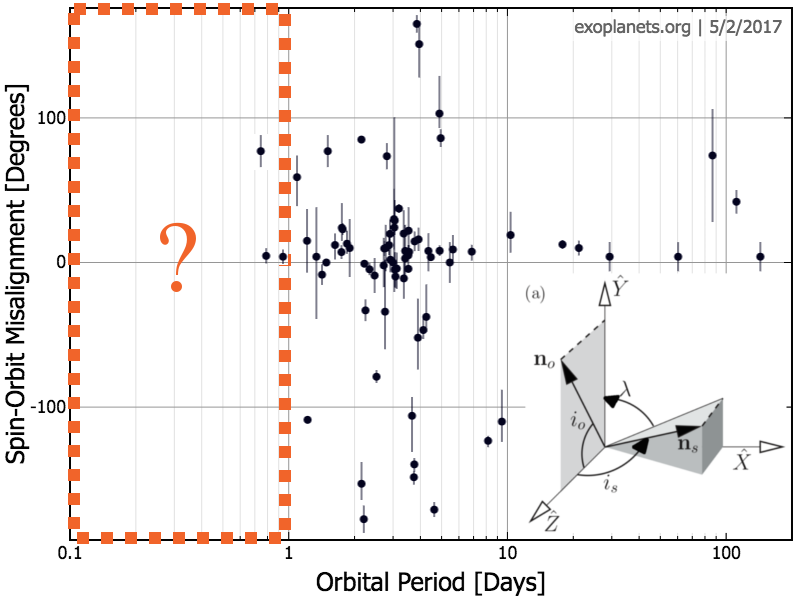}
\caption{Stellar obliquities help distinguish between different origin scenarios, but the spin-orbit alignments for the ultra-short-period targets ($P\le1$~day) are largely unknown. \label{fig:obliquity}}
\end{wrapfigure}

The obliquities of short-period planet-hosting stars may help distinguish between origin scenarios. Originally, large obliquities were thought to point to scenario (2), such as for the HAT-P-7 system, a hot Jupiter in a system with $\lambda = (155 \pm 37)^\circ$ \citep{2012ApJ...757...18A}. Small obliquities were interpreted as evidence for scenario (3) \citep[e.g.][]{1996Natur.380..606L}, consistent with tightly packed, multi-planet systems such as Kepler-30 \citep{2006Sci...313.1413R}. However, mounting evidence has complicated this simple connection: hotter stars tend to have a broader range of obliquities, and \citet{2012ApJ...757...18A} argued this trend indicates close-in planets originated via scenario (2) but that tides preferentially reduced initially large obliquities for the cooler stars, whose tidal dissipation is thought to be more effective \citep{2009ApJ...704..930P}. On the other hand, \citet{2012Natur.491..418B} suggested that a distant stellar companion can warp a protoplanetary disk, possibly producing large obliquities even for planets brought close-in through gas disk migration. 

Small USPs may uniquely contribute to this discussion since their small masses mean that tidal interactions with their host stars probably drive little to no orbital or obliquity evolution, and so  the current obliquities for USP-hosting stars are probably little modified from their original values. Unfortunately, the leading way to estimate obliquity, measuring the Rossiter-McLaughlin effect during transit \citep[e.g.,][]{2005ApJ...631.1215W}, has almost exclusively been measured for planets Saturn-sized or larger\footnote{See John Southworth's TEPCat catalog at \href{http://www.astro.keele.ac.uk/jkt/tepcat/rossiter.html}{www.astro.keele.ac.uk/jkt/tepcat/rossiter.html}.}. Many planets, including almost all USPs, are too small or orbit stars too faint for this technique. The 8 Earth-mass, 1.9 Earth-radius planet 55\,Cnc\,e represents a fascinating exceptional case with hints to a significant orbital misalignment with respect to its host star \citep{2014A&A...569A..65B,2014ApJ...792L..31L}.

Obliquities of USP-hosting stars may be determinable with the starspot-crossing technique \citep[e.g.][]{2011ApJ...740...33D}. Rotational modulation can be used to infer the rotation period \citep{2012A&A...539A.137M}, and if the spots pass through the planet's transit corridor (the portion of the stellar disk occulted during transit), a short (several minutes) spike in the transit curve can result. The frequency and timing of such spikes have previously been used to estimate stellar obliquities \citep{2012Natur.487..449S}, and the high time-resolution provided by the upcoming TESS mission will likely make the data well-suited to search for and analyze such signals for any USPs TESS finds.


\begin{multicols}{2}
\small
\setlength{\bibsep}{0pt}
\bibliography{our_bibliography}

\begin{thebibliography}{}
\expandafter\ifx\csname natexlab\endcsname\relax\def\natexlab#1{#1}\fi

\bibitem[{{Adams} {et~al.}(2016){Adams}, {Jackson}, \&
  {Endl}}]{2016AJ....152...47A}
{Adams}, E.~R., {Jackson}, B., \& {Endl}, M. 2016, \aj, 152, 47

\bibitem[{{Adams} {et~al.}(2017){Adams}, {Jackson}, {Endl}, {Cochran},
  {MacQueen}, {Duev}, {Jensen-Clem}, {Salama}, {Ziegler}, {Baranec},
  {Kulkarni}, {Law}, \& {Riddle}}]{2017AJ....153...82A}
{Adams}, E.~R., {Jackson}, B., {Endl}, M., {et~al.} 2017, \aj, 153, 82

\bibitem[{{Albrecht} {et~al.}(2012){Albrecht}, {Winn}, {Johnson}, {Howard},
  {Marcy}, {Butler}, {Arriagada}, {Crane}, {Shectman}, {Thompson}, {Hirano},
  {Bakos}, \& {Hartman}}]{2012ApJ...757...18A}
{Albrecht}, S., {Winn}, J.~N., {Johnson}, J.~A., {et~al.} 2012, ApJ, 757, 18

\bibitem[{{Batygin}(2012)}]{2012Natur.491..418B}
{Batygin}, K. 2012, \nat, 491, 418

\bibitem[{{Batygin} {et~al.}(2016){Batygin}, {Bodenheimer}, \&
  {Laughlin}}]{2016ApJ...829..114B}
{Batygin}, K., {Bodenheimer}, P.~H., \& {Laughlin}, G.~P. 2016, \apj, 829, 114

\bibitem[{{Bodenheimer} {et~al.}(2000){Bodenheimer}, {Hubickyj}, \&
  {Lissauer}}]{2000Icar..143....2B}
{Bodenheimer}, P., {Hubickyj}, O., \& {Lissauer}, J.~J. 2000, \icarus, 143, 2

\bibitem[{{Bolmont} \& {Mathis}(2016)}]{2016CeMDA.126..275B}
{Bolmont}, E., \& {Mathis}, S. 2016, Celestial Mechanics and Dynamical
  Astronomy, 126, 275

\bibitem[{{Bourrier} \& {H{\'e}brard}(2014)}]{2014A&A...569A..65B}
{Bourrier}, V., \& {H{\'e}brard}, G. 2014, \aap, 569, A65

\bibitem[{{Bouvier} {et~al.}(2007){Bouvier}, {Alencar}, {Harries},
  {Johns-Krull}, \& {Romanova}}]{2007prpl.conf..479B}
{Bouvier}, J., {Alencar}, S.~H.~P., {Harries}, T.~J., {Johns-Krull}, C.~M., \&
  {Romanova}, M.~M. 2007, Protostars and Planets V, 479

\bibitem[{{Chatterjee} {et~al.}(2008){Chatterjee}, {Ford}, {Matsumura}, \&
  {Rasio}}]{2008ApJ...686..580C}
{Chatterjee}, S., {Ford}, E.~B., {Matsumura}, S., \& {Rasio}, F.~A. 2008, \apj,
  686, 580

\bibitem[{{Dawson} \& {Johnson}(2018)}]{2018arXiv180106117D}
{Dawson}, R.~I., \& {Johnson}, J.~A. 2018, ArXiv e-prints, arXiv:1801.06117

\bibitem[{{Dawson} {et~al.}(2015){Dawson}, {Murray-Clay}, \&
  {Johnson}}]{2015ApJ...798...66D}
{Dawson}, R.~I., {Murray-Clay}, R.~A., \& {Johnson}, J.~A. 2015, \apj, 798, 66

\bibitem[{{Deming} {et~al.}(2011){Deming}, {Sada}, {Jackson}, {Peterson},
  {Agol}, {Knutson}, {Jennings}, {Haase}, \& {Bays}}]{2011ApJ...740...33D}
{Deming}, D., {Sada}, P.~V., {Jackson}, B., {et~al.} 2011, Astrophysical
  Journal, 740, 33

\bibitem[{{Dumusque} {et~al.}(2011){Dumusque}, {Udry}, {Lovis}, {Santos}, \&
  {Monteiro}}]{2011A&A...525A.140D}
{Dumusque}, X., {Udry}, S., {Lovis}, C., {Santos}, N.~C., \& {Monteiro},
  M.~J.~P.~F.~G. 2011, \aap, 525, A140

\bibitem[{{Fabrycky} \& {Tremaine}(2007)}]{2007ApJ...669.1298F}
{Fabrycky}, D., \& {Tremaine}, S. 2007, \apj, 669, 1298

\bibitem[{{Ford} \& {Rasio}(2006)}]{2006ApJ...638L..45F}
{Ford}, E.~B., \& {Rasio}, F.~A. 2006, \apjl, 638, L45

\bibitem[{{Hansen} \& {Zink}(2015)}]{2015MNRAS.450.4505H}
{Hansen}, B.~M.~S., \& {Zink}, J. 2015, \mnras, 450, 4505

\bibitem[{{Heller}(2018)}]{Heller2018}
{Heller}, R. 2018, (in prep.)

\bibitem[{{Howard} {et~al.}(2012){Howard}, {Marcy}, {Bryson}, {Jenkins},
  {Rowe}, {Batalha}, {Borucki}, {Koch}, {Dunham}, {Gautier}, {Van Cleve},
  {Cochran}, {Latham}, {Lissauer}, {Torres}, {Brown}, {Gilliland}, {Buchhave},
  {Caldwell}, {Christensen-Dalsgaard}, {Ciardi}, {Fressin}, {Haas}, {Howell},
  {Kjeldsen}, {Seager}, {Rogers}, {Sasselov}, {Steffen}, {Basri},
  {Charbonneau}, {Christiansen}, {Clarke}, {Dupree}, {Fabrycky}, {Fischer},
  {Ford}, {Fortney}, {Tarter}, {Girouard}, {Holman}, {Johnson}, {Klaus},
  {Machalek}, {Moorhead}, {Morehead}, {Ragozzine}, {Tenenbaum}, {Twicken},
  {Quinn}, {Isaacson}, {Shporer}, {Lucas}, {Walkowicz}, {Welsh}, {Boss},
  {Devore}, {Gould}, {Smith}, {Morris}, {Prsa}, {Morton}, {Still}, {Thompson},
  {Mullally}, {Endl}, \& {MacQueen}}]{2012ApJS..201...15H}
{Howard}, A.~W., {Marcy}, G.~W., {Bryson}, S.~T., {et~al.} 2012, \apjs, 201, 15

\bibitem[{{Howard} {et~al.}(2013){Howard}, {Sanchis-Ojeda}, {Marcy}, {Johnson},
  {Winn}, {Isaacson}, {Fischer}, {Fulton}, {Sinukoff}, \&
  {Fortney}}]{2013Natur.503..381H}
{Howard}, A.~W., {Sanchis-Ojeda}, R., {Marcy}, G.~W., {et~al.} 2013, \nat, 503,
  381

\bibitem[{{Jackson} {et~al.}(2017){Jackson}, {Arras}, {Penev}, {Peacock}, \&
  {Marchant}}]{2017ApJ...835..145J}
{Jackson}, B., {Arras}, P., {Penev}, K., {Peacock}, S., \& {Marchant}, P. 2017,
  \apj, 835, 145

\bibitem[{{Jackson} {et~al.}(2008){Jackson}, {Greenberg}, \&
  {Barnes}}]{2008ApJ...678.1396J}
{Jackson}, B., {Greenberg}, R., \& {Barnes}, R. 2008, Astrophysical Journal,
  678, 1396

\bibitem[{{Jackson} {et~al.}(2013){Jackson}, {Stark}, {Adams}, {Chambers}, \&
  {Deming}}]{2013ApJ...779..165J}
{Jackson}, B., {Stark}, C.~C., {Adams}, E.~R., {Chambers}, J., \& {Deming}, D.
  2013, \apj, 779, 165

\bibitem[{{Jia} \& {Spruit}(2017)}]{2017MNRAS.465..149J}
{Jia}, S., \& {Spruit}, H.~C. 2017, \mnras, 465, 149

\bibitem[{{Kuchner} \& {Lecar}(2002)}]{2002ApJ...574L..87K}
{Kuchner}, M.~J., \& {Lecar}, M. 2002, \apjl, 574, L87

\bibitem[{{Lee} \& {Chiang}(2017)}]{2017ApJ...842...40L}
{Lee}, E.~J., \& {Chiang}, E. 2017, \apj, 842, 40

\bibitem[{{Lesniak} \& {Desch}(2011)}]{2011ApJ...740..118L}
{Lesniak}, M.~V., \& {Desch}, S.~J. 2011, \apj, 740, 118

\bibitem[{{Lin} {et~al.}(1996){Lin}, {Bodenheimer}, \&
  {Richardson}}]{1996Natur.380..606L}
{Lin}, D.~N.~C., {Bodenheimer}, P., \& {Richardson}, D.~C. 1996, \nat, 380, 606

\bibitem[{{Lodders}(2003)}]{2003ApJ...591.1220L}
{Lodders}, K. 2003, \apj, 591, 1220

\bibitem[{{L{\'o}pez-Morales} {et~al.}(2014){L{\'o}pez-Morales}, {Triaud},
  {Rodler}, {Dumusque}, {Buchhave}, {Harutyunyan}, {Hoyer}, {Alonso}, {Gillon},
  {Kaib}, {Latham}, {Lovis}, {Pepe}, {Queloz}, {Raymond}, {S{\'e}gransan},
  {Waldmann}, \& {Udry}}]{2014ApJ...792L..31L}
{L{\'o}pez-Morales}, M., {Triaud}, A.~H.~M.~J., {Rodler}, F., {et~al.} 2014,
  \apjl, 792, L31

\bibitem[{{McQuillan} {et~al.}(2012){McQuillan}, {Aigrain}, \&
  {Roberts}}]{2012A&A...539A.137M}
{McQuillan}, A., {Aigrain}, S., \& {Roberts}, S. 2012, \aap, 539, A137

\bibitem[{{Metzger} {et~al.}(2012){Metzger}, {Giannios}, \&
  {Spiegel}}]{2012MNRAS.425.2778M}
{Metzger}, B.~D., {Giannios}, D., \& {Spiegel}, D.~S. 2012, \mnras, 425, 2778

\bibitem[{{Penev} {et~al.}(2009){Penev}, {Sasselov}, {Robinson}, \&
  {Demarque}}]{2009ApJ...704..930P}
{Penev}, K., {Sasselov}, D., {Robinson}, F., \& {Demarque}, P. 2009, \apj, 704,
  930

\bibitem[{{Plavchan} \& {Bilinski}(2013)}]{2013ApJ...769...86P}
{Plavchan}, P., \& {Bilinski}, C. 2013, \apj, 769, 86

\bibitem[{{Raymond} {et~al.}(2006){Raymond}, {Mandell}, \&
  {Sigurdsson}}]{2006Sci...313.1413R}
{Raymond}, S.~N., {Mandell}, A.~M., \& {Sigurdsson}, S. 2006, Science, 313,
  1413

\bibitem[{{Sanchis-Ojeda} {et~al.}(2014){Sanchis-Ojeda}, {Rappaport}, {Winn},
  {Kotson}, {Levine}, \& {El Mellah}}]{2014ApJ...787...47S}
{Sanchis-Ojeda}, R., {Rappaport}, S., {Winn}, J.~N., {et~al.} 2014, \apj, 787,
  47

\bibitem[{{Sanchis-Ojeda} {et~al.}(2013){Sanchis-Ojeda}, {Rappaport}, {Winn},
  {Levine}, {Kotson}, {Latham}, \& {Buchhave}}]{2013ApJ...774...54S}
---. 2013, \apj, 774, 54

\bibitem[{{Sanchis-Ojeda} {et~al.}(2012){Sanchis-Ojeda}, {Fabrycky}, {Winn},
  {Barclay}, {Clarke}, {Ford}, {Fortney}, {Geary}, {Holman}, {Howard},
  {Jenkins}, {Koch}, {Lissauer}, {Marcy}, {Mullally}, {Ragozzine}, {Seader},
  {Still}, \& {Thompson}}]{2012Natur.487..449S}
{Sanchis-Ojeda}, R., {Fabrycky}, D.~C., {Winn}, J.~N., {et~al.} 2012, \nat,
  487, 449

\bibitem[{{Spalding} \& {Batygin}(2015)}]{2015ApJ...811...82S}
{Spalding}, C., \& {Batygin}, K. 2015, \apj, 811, 82

\bibitem[{{Sullivan} {et~al.}(2015){Sullivan}, {Winn}, {Berta-Thompson},
  {Charbonneau}, {Deming}, {Dressing}, {Latham}, {Levine}, {McCullough},
  {Morton}, {Ricker}, {Vanderspek}, \& {Woods}}]{2015ApJ...809...77S}
{Sullivan}, P.~W., {Winn}, J.~N., {Berta-Thompson}, Z.~K., {et~al.} 2015, \apj,
  809, 77

\bibitem[{{Terquem}(2003)}]{2003MNRAS.341.1157T}
{Terquem}, C.~E.~J.~M.~L.~J. 2003, \mnras, 341, 1157

\bibitem[{{Trilling} {et~al.}(1998){Trilling}, {Benz}, {Guillot}, {Lunine},
  {Hubbard}, \& {Burrows}}]{1998ApJ...500..428T}
{Trilling}, D.~E., {Benz}, W., {Guillot}, T., {et~al.} 1998, \apj, 500, 428

\bibitem[{{Valsecchi} {et~al.}(2015){Valsecchi}, {Rappaport}, {Rasio},
  {Marchant}, \& {Rogers}}]{2015ApJ...813..101V}
{Valsecchi}, F., {Rappaport}, S., {Rasio}, F.~A., {Marchant}, P., \& {Rogers},
  L.~A. 2015, \apj, 813, 101

\bibitem[{{Winn} {et~al.}(2010){Winn}, {Fabrycky}, {Albrecht}, \&
  {Johnson}}]{2010ApJ...718L.145W}
{Winn}, J.~N., {Fabrycky}, D., {Albrecht}, S., \& {Johnson}, J.~A. 2010, \apjl,
  718, L145

\bibitem[{{Winn} {et~al.}(2005){Winn}, {Noyes}, {Holman}, {Charbonneau},
  {Ohta}, {Taruya}, {Suto}, {Narita}, {Turner}, {Johnson}, {Marcy}, {Butler},
  \& {Vogt}}]{2005ApJ...631.1215W}
{Winn}, J.~N., {Noyes}, R.~W., {Holman}, M.~J., {et~al.} 2005, \apj, 631, 1215

\end{thebibliography}
\bibliographystyle{apj}
\end{multicols}

\end{document}